\documentclass[twocolumn]{aastex7}

\usepackage{subcaption}
\usepackage{multirow}
\DeclareUnicodeCharacter{00A0}{~}
\begin{document}

\title{Evidence for Distinct One-Off and Repeating Subpopulations in CHIME/FRB Catalog 2}

\correspondingauthor{Sloane Sirota}
\email{ss00300@mix.wvu.edu}

\author[orcid=0009-0008-6718-172X,gname=Sloane, sname='Sirota']{Sloane Sirota} 
\affiliation{Department of Physics and Astronomy,  West Virginia University, Morgantown, WV 26506-6315, USA}
\affiliation{Center for Gravitational Waves and Cosmology, Chestnut Ridge Building, Morgantown, WV 26505-6315, USA}
\email{ss003003@mix.wvu.edu}

\author[orcid=0000-0003-1301-966X,gname=Duncan,sname='Lorimer']{Duncan R. Lorimer} 
\affiliation{Department of Physics and Astronomy,  West Virginia University, Morgantown, WV 26506-6315, USA}
\affiliation{Center for Gravitational Waves and Cosmology, Chestnut Ridge Building, Morgantown, WV 26505-6315, USA}
\affiliation{National Centre for Radio Astrophysics, Tata Institute of Fundamental Research, Pune, 410007, India}
\email{duncan.lorimer@mail.wvu.edu}

\begin{abstract}
The Canadian Hydrogen Intensity Mapping Experiment (CHIME) has published its second catalog of 3,641 fast radio bursts (FRBs). While this sample includes 981 bursts from 83 repeating sources, only 2.3\% of CHIME FRBs are observed to repeat. This raises the question of whether the FRB population  truly consists of a single repeating population or includes a distinct subpopulation of one-off sources. We apply Simulation Based Inference (SBI), a likelihood-free machine learning framework, to model the CHIME/FRB Catalog 2 dataset, testing eight FRB population models, half of which include a subpopulation of truly one-off sources. The current catalog strongly and decisively favors models including such a subpopulation, with the best-fitting model combining a log-normal burst rate distribution, a Poissonian wait-time distribution, and a one-off subpopulation. This preference is corroborated by an independent random forest classifier and by the ability of one-off models to naturally reproduce the hyperactive repeating sources present in the catalog without having been trained on them.
We place upper limits on the potentially observable CHIME/FRB population of 36,000 repeating sources and 19 one-off sources per day in the CHIME sky. Models with a one-off subpopulation naturally account for the hyperactive repeating sources present in the catalog without having been trained on them, providing an independent consistency check on our results. We predict the repeater percentage will fall below 2\% in the next five years, eventually declining to $\sim$0.5\% as the sample saturates with one-off detections.
\end{abstract}


\section{Introduction} 

Fast radio bursts (FRBs) are energetic ($L \sim 10^{36-44}$ erg/s) pulses of radio emission detected at cosmological distances \citep{duncburst,2013Sci...341...53T}. FRBs currently fall into two broad categories: apparent one-off and, following the discovery of multiple pulses from
FRB~20121102A \citep{first_repeating_frb}, repeating sources. Whether or not these categories are a result of observational biases or reflective of distinct physical differences is currently unclear \citep{2020MNRAS.497.3076C,clancy_frbs_can_repeat,frb_repeater_nonrepeater_energy_func,Kharel_2026}.

The rapidly growing sample of FRBs leads us to ask whether a solely repeating population is sufficient to explain all observed sources. Previous efforts to constrain the repeater fraction remain inconclusive \citep[e.g.,][]{amandas_repeater_paper,yamasaki2023truefractionrepeatingfast}, even as non-cataclysmic progenitor models continue to predict that all sources eventually repeat \citep[e.g.,][]{Wadiasingh_2019,paz_frbs_can_repeat}. We use novel methods to determine whether current observations align with a unified repeating population of FRBs.

In this paper we analyze a large sample of FRBs discovered by the Canadian Hydrogen Intensity Mapping Experiment (CHIME) telescope. CHIME observes the sky north of declination $\delta > -11^{\circ}$ with daily cadence in the 400--800 MHz frequency band. Its unprecedented exposure led to the first CHIME/FRB Catalog \citep[hereafter Catalog 1,][]{chimecat1}, which reported 536 bursts from 492 unique sources (474 apparent one-off sources and 18 repeating sources), expanding the known FRB population by an order of magnitude. The second CHIME/FRB Catalog \citep[hereafter Catalog 2,][]{chimecat2} significantly increases this sample, reporting 4,539 total bursts from 3,641 unique sources (3,558 apparent one-off sources and 83 repeating sources).

Our initial attempts to model this population encountered a large amount of stochasticity making it non-trivial to determine a likelihood function via classical Markov-Chain Monte Carlo simulations. Progress in tackling such problems use an emerging technique known as simulation based inference (SBI), a machine learning method designed to model statistical problems such as this one where closed-form expressions for the likelihood function do not exist \citep{sbi_2025}. SBI was recently applied successfully to study the Galactic population of magnetars 
\citep{Sautron_2025}.

We use SBI to test and compare a set of population models for FRBs and determine if they can properly predict the evolution of the CHIME/FRB population between Catalog 1 and Catalog 2. In Section \ref{sec:meth}, we describe our modeling approach in detail. In Section \ref{sec:models}, we discuss the population models that we test. We present our results and discuss their implications  and future predictions in Section \ref{sec:resdisc}. We draw  conclusions and suggest future work in Section \ref{sec:conc}.

\section{Methods}\label{sec:meth}
We explore models of the FRB population visible to CHIME using SBI techniques. Our overarching approach is to assume that there are a finite number of FRBs in the CHIME sky that are potentially detectable and that all of them repeat. As detailed below, we also consider models that contain a fraction of one-off bursts. 

\subsection{CHIME/FRB sample}\label{chimesample}
Following \cite{yamasaki2023truefractionrepeatingfast}, we exclude detections from the first six months of the CHIME observing window due to fluctuations in sensitivity and exposure in this period \citep{chimecat1}. We also exclude all side-lobe FRB detections which do not have logged header localizations \citep{chimecat2}. 
To avoid any bias, we remove the outlying hyperactive sources FRB20220912A and FRB20180916B. We comment on these sources in the context of our models in Section \ref{sec:hyper}. After these omissions, we end up with a sample of 3,854 FRBs out of the 4,539 FRBs from Catalog 2 observed between 2018, December 31 and 2023, September 15. 
 
\subsection{The CHIME/FRB selection function}
The 3,854 FRBs represents an observationally selected sample that is a subset of all potentially observable FRBs in the CHIME sky. By potentially observable, we mean all FRBs with fluences above the CHIME/FRB detection threshold at the instrument's meridian. We consider a  population where 
there are a finite number of repeating FRBs which we seek to constrain
along with models containing a rate of potentially observable one-off FRBs.
We choose to model both the CHIME/FRB sensitivity and exposure (both of which vary with declination) in one representative {\it selection function}, i.e.~the probability of detecting
a model FRB during the SBI runs. This approach is a simplified version of the work of \cite{kyle_selection_function} which was not available to us at the time of developing our analysis. 

\begin{figure}
    \centering
    \includegraphics[width=\linewidth]{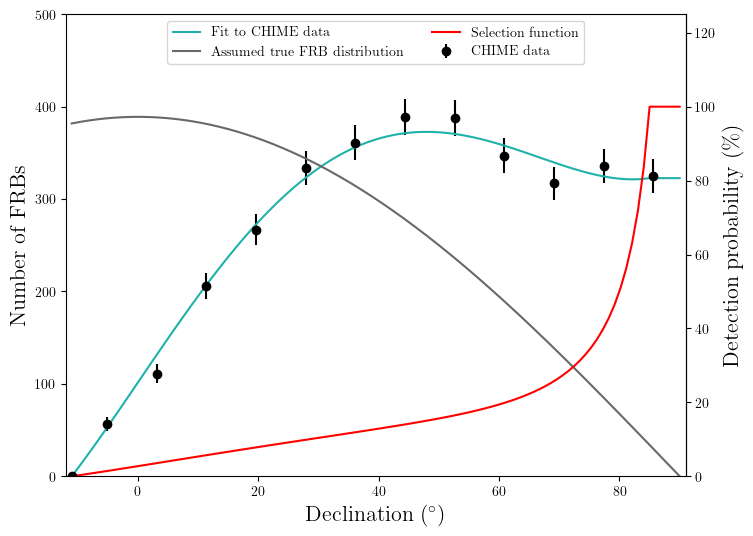}
    \caption{The number of CHIME/FRB detections in Catalog 2 as a function of declination, plotted with the expected true distribution, and the fitted selection function.\label{fig:selection_func}}
\end{figure}

We derive the selection function by making the reasonable assumption that FRBs are uniformly distributed across the sky \citep{frb_all_sky_rate}. As a result, the underlying probability density function of FRBs as a function of declination $\delta$ scales as $\cos \delta$. Our  model of the CHIME/FRB selection function,
$P_{\rm selection}(\delta)$, satisfies the condition that the observed number of FRBs
\begin{equation}\label{selfunc}
  N_{\rm FRBs,Observed}(\delta) = P_{\rm selection}(\delta) \times N_{\rm FRBs,True}(\delta), 
\end{equation}
where $N_{\rm FRBs,True}(\delta) \propto \cos \delta$ is the true distribution of FRBs.
To get $N_{\rm FRBs,Observed}(\delta)$, we binned the observed distribution of FRB declinations into 13 bins and fit a smooth curve to the data. We tested polynomials of orders three through eight, to see what order could optimize the reduced $\chi^2$ value and minimize our parameter uncertainties. We also tested combinations of polynomials where we set certain coefficients to zero. The optimal polynomial order is a fourth-order polynomial with the second-order coefficient set to zero, which yields a reduced $\chi^2 \sim 1.2$. The best-fit parameters to the CHIME/FRB Catalog 2 distribution are shown in Table~\ref{tab:polyfit}. We add a bin of height zero at a declination of $-11^{\circ}$ to ensure that our fit correctly models the CHIME/FRB observation region. We only use the first detection of each repeating FRB source in order to avoid biases from repeating FRBs in our simplified selection function. The polynomial fit can be seen in Fig.~\ref{fig:selection_func}, where the fitted curve is plotted with the assumed true FRB distribution in the sky as a function of declination.

\begin{table}[h]
\centering
\caption{Coefficients of the 4th-order polynomial fit $f(x) = \sum_{i=0}^4 a_i x^i$ to the CHIME/FRB detections as a function of declination. Uncertainties represent the 1$\sigma$ errors in the fitted coefficients. Since it is not found to be statistically significant, we set the second-order coefficient $a_2=0$.}
\begin{tabular}{c c}
\decimals
\hline
Coefficient & Value $\pm$ 1$\sigma$ \\
\hline
$a_0$ & $(10.6 \pm 0.3) \times 10^{1}$ \\
$a_1$ & $10.1 \pm 0.3$ \\
$a_3$ & $(-2.8 \pm 0.3) \times 10^{-3}$ \\
$a_4$ & $(2.1 \pm 0.4) \times 10^{-5}$ \\
\hline
\end{tabular}
\label{tab:polyfit}
\end{table}

 Using our fitted polynomial curve for $N_{\rm FRBs,Observed}(\delta)$ in Eqn.~\ref{selfunc}, we then obtain a selection function pictured in Fig.~\ref{fig:selection_func}. Due to the behavior
 of the cosine function, $N_{\rm FRBs,True}(\delta) \rightarrow 0$ as $\delta \rightarrow 90^{\circ}$. We therefore truncate the selection function once it reaches unity, as shown in Fig.~\ref{fig:selection_func}. We then scale our selection function so its maximum is 0.99 to reflect that there is never a 100\% chance of detecting an FRB due to various factors including weather and telescope maintenance. In this framework, the selection function reflects the detectability of FRBs in the CHIME sky with the current instrument which is a function of both instrumental
 sensitivity and exposure time as a function of declination. Throughout the analyses performed in this paper, we use this function to model the detectability of a simulated FRB with declination $\delta$ by sampling a uniform random deviate $U$ between zero and one. Such FRBs are deemed detectable when $U<P_{\rm selection}(\delta)$. The selection function shape, seen in Fig.~\ref{fig:selection_func}, reflects mostly the exposure time of the CHIME telescope because exposure affects the detection probability more than sensitivity as a function of declination \citep{chime_operations}.

\subsection{Simulation based inference}
\begin{figure}
    \centering
    \includegraphics[width=0.95\linewidth]{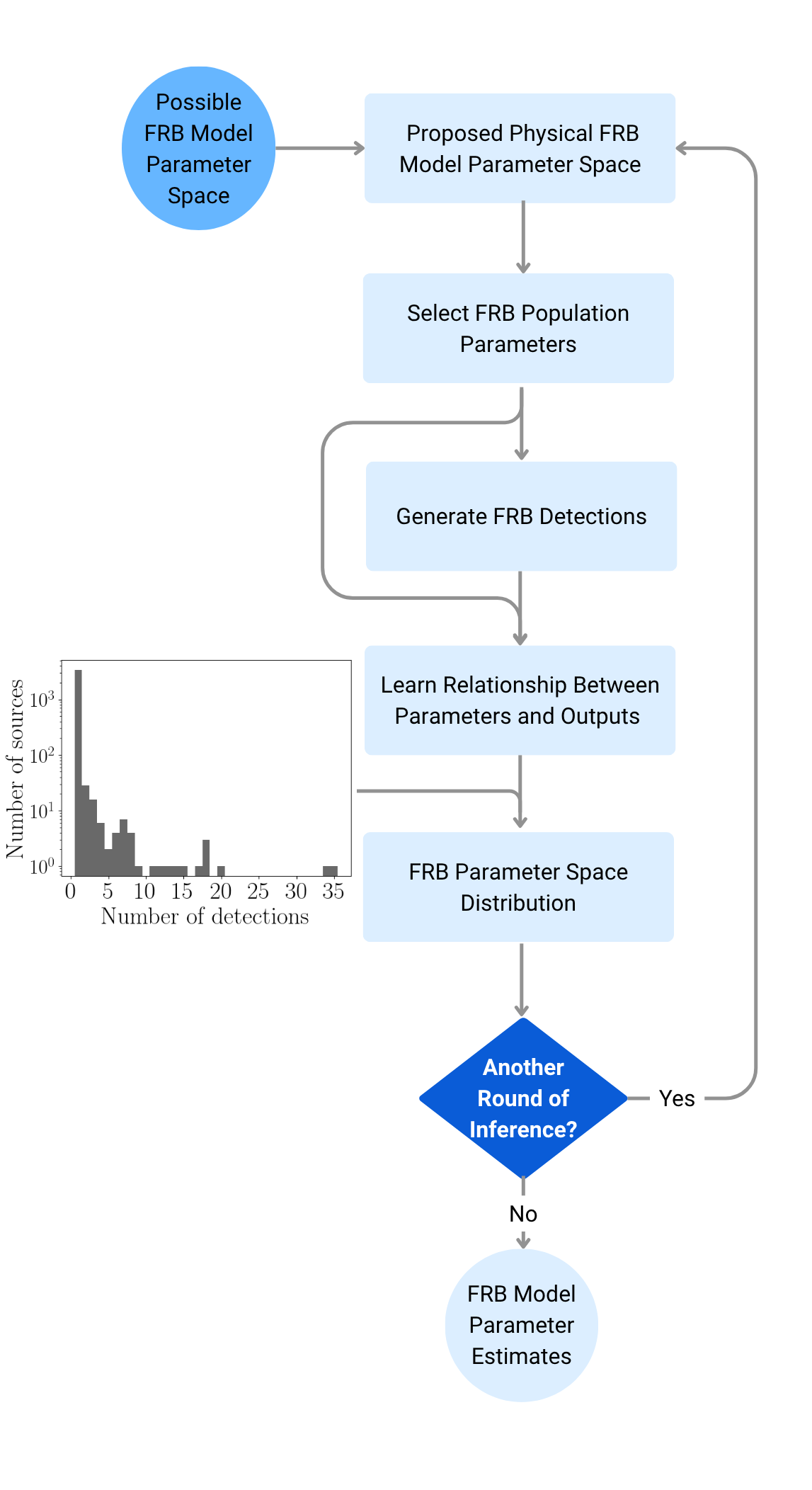}
    \caption{The SBI methodology as applied to this work. We input the possible parameter space into our model, which  is used as our proposal for the first round of inference. Each simulation iteration selects a set of FRB population parameters within the parameter ranges. These population parameters  run through the simulation, which generates a simulated FRB population that aims to mimic the histogram of the number of bursts per source shown in the inset. Both the input parameters and the output FRB population are then fed into the neural network so it can learn the relationship between the parameters and the outputs. We then take the Catalog 2 observations as inputs to the neural network, so that it can output estimates on the posterior distributions. We then repeat this cycle with the posterior estimates as our new proposals and with a specific observation we aim to model in what is called a sequential process. We repeat this, typically for 3--5 rounds, until the model is well-trained on the desired observation and the posterior distributions stop changing between rounds.}
    \label{fig:sbi_flow}
\end{figure}

Closed-form likelihood functions are ideally required when carrying out Markov Chain Monte Carlo simulations of physical systems in an attempt to determine posterior distributions and parameter estimates. Complications arise when using simulations with inherent stochasticity to determine the likelihood function. Indeed, our initial efforts in this regard were unsuccessful due to unstable
estimates of the likelihood function. Simulation based inference (SBI) offers a novel solution to such problems and is the approach we take in this study. SBI utilizes a neural network to learn the relationship between model input parameters and output parameter distributions to accurately characterize the true posterior parameter distributions. In our case, we aim to model the number of Catalog 2 sources with a given number of observed bursts, which quantifies the observed FRB repeater rates. We implement this using the \texttt{SBI} Python package \citep{sbi_2025}. For this work, where we are mainly concerned with  mapping the number of FRB detections to the underlying population parameters, we utilize a sequential neural likelihood estimator \citep[SNLE;][]{snle}. This framework is shown in Fig.~\ref{fig:sbi_flow}. To determine the validity of our results, we utilize local classifier two-sample tests, which checks the calibration of the SNLEs at a given observation \citep{linhart2023c2st}. 

We then compare our models using the learnt harmonic mean estimator, developed and implemented using the Python package \texttt{harmonic} \citep{harmonic,spurio-mancini:harmonic_sbi}. We train the estimators using a rational quadratic spline, a type of normalizing flow, to accurately obtain the Bayesian evidence of our models using the marginal likelihoods derived from our sequential neural likelihood estimator \citep{normalizing_flow_harmonic,spurio-mancini:harmonic_sbi}. 

As a secondary check of the learnt harmonic mean estimator, we train a random forest regression classifier on the model priors to choose the favored population model using the \texttt{scikit-learn} Python software \citep{scikit-learn}. The random forest is made up of decision trees, each of which tests random features of the models against our data point. In our case, these features are histogram bins. After the decision trees learn their random set of histogram bins for the different models, we feed in the data point we are trying to fit for. In this framework, this is the CHIME Catalog 2 data. After being fed these data, the decision trees effectively ``vote'' for which model is likely to have produced this data based on their training. The favored model is the one with the most votes from different decision trees.

\section{FRB Population Models}\label{sec:models}

We test all combinations of burst rate and wait time distribution models shown in Table \ref{table:priors}. Each model attempts to match the histogram of the number of CHIME/FRB sources with a given number of detections shown in the inset of Fig.~\ref{fig:sbi_flow}. We then see if this time-independently trained model can match the time evolution of Catalog 2 for the overall number of sources, the number of repeating sources and the number of detections from repeating sources.
Because we are exploring eight different models throughout this work, we adopt an abbreviated naming convention to refer to them throughout the paper with up to three letters denoting each model. The details of the scheme are given below.

\begin{deluxetable*}{l | c c c | c c | c | c | c}[ht]
\tablecolumns{9}
\tablecaption{Prior ranges used in our SBI analysis. From left to right, we list the base 10 logarithms of the power-law burst rate range ($r_{\rm min}$ and $r_{\rm max}$ in units of bursts~hr$^{-1}$) and exponent $q$, the base-10 logarithms of the log-normal burst rate distribution ($\mu$ and $\sigma$ in units of $\ln$ bursts~hr$^{-1}$), the total number of repeating FRBs ($N$), the Weibull wait-time distribution shape parameter ($k$) and the base-10 logarithm of the rate of one-off FRBs in the CHIME/FRB sky ($r_{\rm oo}$ in units of FRBs~day$^{-1}$). \label{table:priors}}
\tabletypesize{\scriptsize}
\tablewidth{\textwidth}
\decimals
\tablehead{Model     & \multicolumn{3}{l|}{Power-Law (P)}               & \multicolumn{2}{l|}{Log-normal (L)}    & All         & Weibull (W) & One-Offs (O)}
\startdata
Parameter & $\log_{10}(r_{\rm min})$ & $\log_{10}(r_{\rm max})$ & $q$ & $\log_{10}(\mu)$ & $\log_{10}(\sigma)$ & $\log_{10}(N)$ & $k$         & $\log_{10}(r_{\rm oo})$     \\
\hline
Prior minimum & --5.5 & --1.0 & 2.0 & --9.0 & 0.0 & 3.0 & 0.2 & 0  \\
Prior maximum & --2.0 & 1.0  & 5.5 & --3.0 & 3.5 & 6.0 & 0.9 & 3 \\
\enddata
\end{deluxetable*}
\subsection{Burst Rate Modeling}

\subsubsection{Power-Law Burst Rate}
For this model, we generate FRBs from a repetition rate distribution of the form 
\begin{equation}\label{rep_rate_power_law}
    \frac{{\rm d}N}{{\rm d}r} \propto r^{-q},
\end{equation}
where $N$ is the number of FRBs with a given repetition rate, $r$ is the repetition rate (bursts/hr), and $q > 0$ is a power law index. A power-law drop off could account for the number of highly active repeating FRBs that has been observed, as compared to less active sources and thus far one-off sources. We fit for $r_{\rm min}$ and $r_{\rm max}$, the minimum and maximum repetition rate, respectively. We choose the prior parameter ranges based on those tested in \cite{kyle_dunc_paper}, \cite{yamasaki2023truefractionrepeatingfast} and  \cite{modelling_FRB_pop}. Our prior ranges for $q$ and $N$ are also informed by literature predictions \citep{yamasaki2023truefractionrepeatingfast}. The prior ranges used can be found in Table \ref{table:priors}, where it is shown that we explore the parameter space  of our repetition rate bounds and the number of FRBs in log space. All models using the power-law burst rate model will be referred to with an initial letter `P' (e.g., Model P). 

\subsubsection{Lognormal Burst Rate}
Another common way to model the burst rate of FRB sources is by sampling from a log-normal distribution. We draw repetition rates $r$ from a distribution of the form
\begin{equation}
    \frac{{\rm d}N}{{\rm d}r} \propto \frac{1}{r} \exp\left[ -\frac{(\ln r - \mu)^2}{2\sigma^2} \right],
\end{equation}
where $\ln r$ is normally distributed with mean $\mu$ and standard deviation $\sigma$. A log-normal distribution heavily weighted toward low burst rates could naturally account for the small fraction of observed repeaters. Although the rate model is defined in natural logarithmic space, we parameterize our priors in $\log_{10}$ space (see Table~\ref{table:priors}). All models using the lognormal burst rate model will be referred to by a model name beginning with L, making our two branches of models those that start with P and those that start with L.

\subsection{Burst Wait-time Modeling}
\subsubsection{Poissonian Wait-Time}
For a Poissonian wait-time distribution, all bursts from each individual source are time-independent. We take the average number of bursts expected for each source during the Catalog 2 timeline, then we sample from a Poissonian distribution to determine the true number of bursts we might detect from the simulated source during our observation timeline. Models using the Poissonian wait-time distribution will be denoted by a P after the burst rate model letter (e.g.,~Model PP or Model LP). 

\subsubsection{Weibullian Wait-Time}
Some FRBs have exhibited burst memory, where the probability of seeing a burst changes with the amount of time since the previous burst \citep{Wang_2023,FRB20180916D}. A Weibull distribution has a probability density of the form 
\begin{equation}
    P(x) = \left\{
  \begin{array}{lr} 
      \frac{k}{\lambda}(\frac{x}{\lambda})^{k-1}e^{-(x/\lambda)^k}& x\geq 0 \\
      0 & x < 0 
      \end{array},
\right.
\end{equation}
where $x$ is the time between bursts, $k$ is the shape parameter and $\lambda$ is the size parameter. These parameters are related to the mean of the distribution 
\begin{equation}
    \mu = \lambda \Gamma(1+1/k),
\end{equation}
where 
$\Gamma (x)=\int _{0}^{\infty }t^{x-1}e^{-t}\,{\rm d}t$.
The Weibull wait-time distribution has been used for burst modeling of multiple hyperactive repeating FRB sources, with previous fits informing our parameter choice ranges shown in Table \ref{table:priors} \citep{FRB20180916D,Oppermann_2018}. All models using a Weibull wait-time distribution will be denoted by a W after the burst rate model letter. 

\subsection{A subpopulation of truly one-off sources}
For each combination of burst rate and wait time, we perform two simulations: one which is a solely repeating FRB population and one that has a subpopulation of truly one-off FRBs. We model the one-off subpopulation by fitting for the mean number of one-off bursts per day, $r_{\rm oo}$, where the number of one-offs per day is a Poisson process. The parameter $\log_{10}{N}$ represents the total number of repeating FRB sources, rather than the total number of FRBs. Models that fit for one-offs will be indicated by an O following the wait time distribution indicator (e.g., Model PWO).

\begin{figure}[h!]
    \centering
    \includegraphics[width=\linewidth]{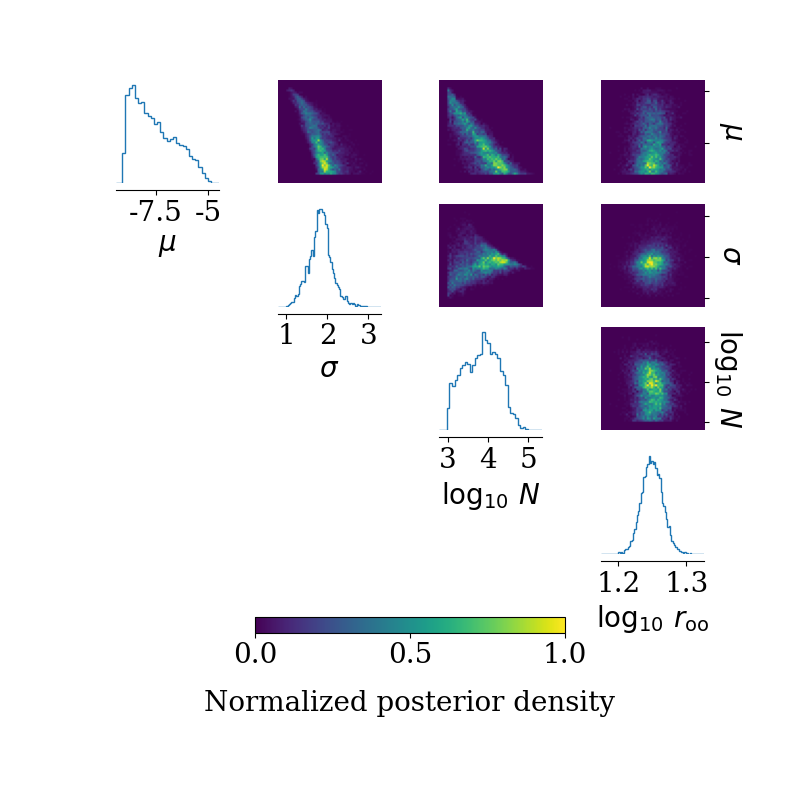}
    \caption{Corner plots showing the posterior distributions for the LPO model.}
    \label{fig:LPO_posterior}
\end{figure}

\section{Results and Discussion}\label{sec:resdisc}

\begin{figure}[h!]
    \centering
    \includegraphics[width=\linewidth]{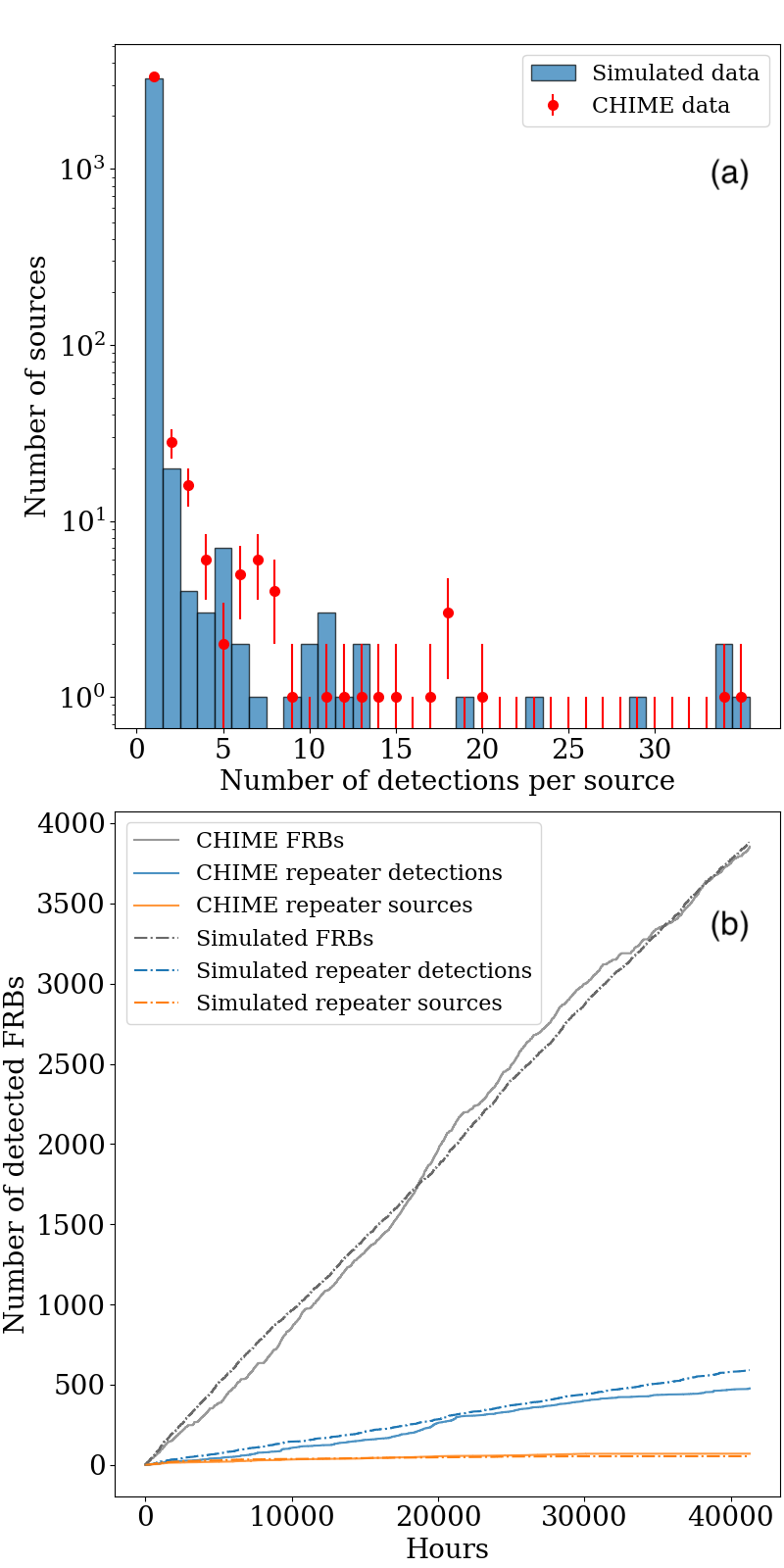}
    \caption{Results from LPO, our best model: (a)  number of sources with a given number of detections compared to Catalog 2 where the error bars assume Poissonian statistics; (b)  time evolution for Catalog 2 compared to LPO showing the total number of FRBs, the number of repeater detections, and the number of repeating sources.}
    \label{fig:summarystatistics}
\end{figure}

Using our SBI-based framework, which attempts to model the Catalog 2 detection statistics as simply as possible, we have
explored a family of eight models for the FRB population. 
The 95\% credible intervals for all model parameters are presented in Table \ref{table:parameterfits}. As discussed below, the statistically favored model is LPO: the lognormal burst rate distribution with a Poissonian wait-time distribution and a subpopulation of one-off sources.  The posterior distributions for this model are shown in Fig.~\ref{fig:LPO_posterior} and its summary statistics (which in this case is a histogram showing the number of pulses per source) compared to Catalog 2 can be seen in Fig.~\ref{fig:summarystatistics}a. 
As shown in Fig.~\ref{fig:summarystatistics}b, LPO provides a good match to the number of repeating and non-repeating FRBs seen in Catalog 2 as a function of time. In the sections below, we
explore the implications of our results.

\begin{deluxetable*}{l c c c c c c c c c c}[ht!]
\tablecolumns{11}
\tablewidth{\textwidth}
\tablecaption{Compilation of Bayes factors and posterior population model parameter 95\% credible intervals for each model. From left to right, we list $\Delta \ln \text{BF}$ (the log Bayes factor relative to the highest evidence model; LPO), the base-10 logarithms of the power-law burst rate range ($r_{\rm min}$ and $r_{\rm max}$ in units of bursts~hr$^{-1}$) and exponent $q$, the base 10 logarithms of the log-normal burst rate distribution ($\mu$ and $\sigma$ in units of $\ln$ bursts~hr$^{-1}$), the total number of repeating FRBs ($N$), the Weibull wait-time distribution shape parameter ($k$) and the base-10 logarithm of the rate of one-off FRBs in the CHIME/FRB sky ($r_{\rm oo}$ in units of FRBs~day$^{-1}$). Finally, as discussed in Section \ref{sec:implications}, we also list the base-10 logarithm of $r_{\rm r}$, the predicted rate of new repeating FRBs in the CHIME sky per month, derived using the fitted quantities of each model.  \label{table:parameterfits}}
\decimals
\tablehead{Model & $\Delta\ln\text{BF}$ & $\log_{10}(r_{\rm min})$ & $\log_{10}(r_{\rm max})$ & $q$ & $\log_{10}(\mu)$ & $\log_{10}(\sigma)$ & $\log_{10}(N)$ & $k$ & $\log_{10}r_{\rm oo}$& $\log_{10} r_{\rm r}$}
\startdata
LPO & 0.00 & - & - & - & $[-9.00, -6.03]$ & $[1.34, 2.44]$ & $[3.09, 4.56]$ & - & $[1.22, 1.28]$ &1.0\\
LWO & 2.87 & - & - & - & $[-8.84,-5.54]$ & $[1.27,2.42]$ & $[3.00,4.07]$ & $[0.24,0.88]$ & $[1.21,1.28]$ & 1.1\\
PPO & 8.49 & $[-5.41,-4.48]$ & $[-0.89,1.00]$ & $[2.00,2.26]$ & - & - & $[3.00,3.82]$ & - & $[1.15,1.28]$ & 1.4\\
PWO & 13.47 & $[-5.50,-4.71]$ & $[-0.90,1.00]$ & $[2.00,2.46]$ & - & - & $[3.00,4.13]$ & $[0.56,0.90]$ & $[1.13,1.30]$ & 1.7 \\
PW & 18.80 & $[-5.10,-4.82]$ & $[-0.99,0.91]$ & $[4.46,5.49]$ & - & - & $[5.11,6.00]$ & $[0.20,0.85]$ & - & 2.8\\
PP & 41.42 & $[-5.50,-5.45]$ & $[-1.00,0.82]$ & $[4.51,5.50]$ & - & - & $[5.20,5.28]$ & - & - &2.7\\
LW & 63.49 & - & - & - & $[-6.42,-4.45]$ & $[0.05, 0.86]$ & $[5.54,6.00]$ & $[0.20,0.77]$ & - & 3.3\\
LP & 114.41 & - & - & - & $[-7.18, -6.93]$ & $[0.87, 0.95]$ & $[5.93, 6.00]$ & - & - &2.6\\
\enddata
\end{deluxetable*}
\subsection{Bayesian evidence}

We compare models using the log Bayes factor, $\Delta\ln\text{BF}$, defined as the difference in Bayesian log-evidence between two models. Differences of $\Delta\ln\text{BF} > 2.3$ and $> 4.6$ represent strong and decisive evidence, respectively, for one model over another \citep{bayesfactor}. As shown in Table \ref{table:parameterfits}, the LPO model is strongly favored over the LWO model ($\Delta\ln\text{BF} = 2.87$) and decisively preferred over all remaining models.

\subsection{Random forest classifier}
\begin{figure*}
    \centering
    \includegraphics[width=\linewidth]{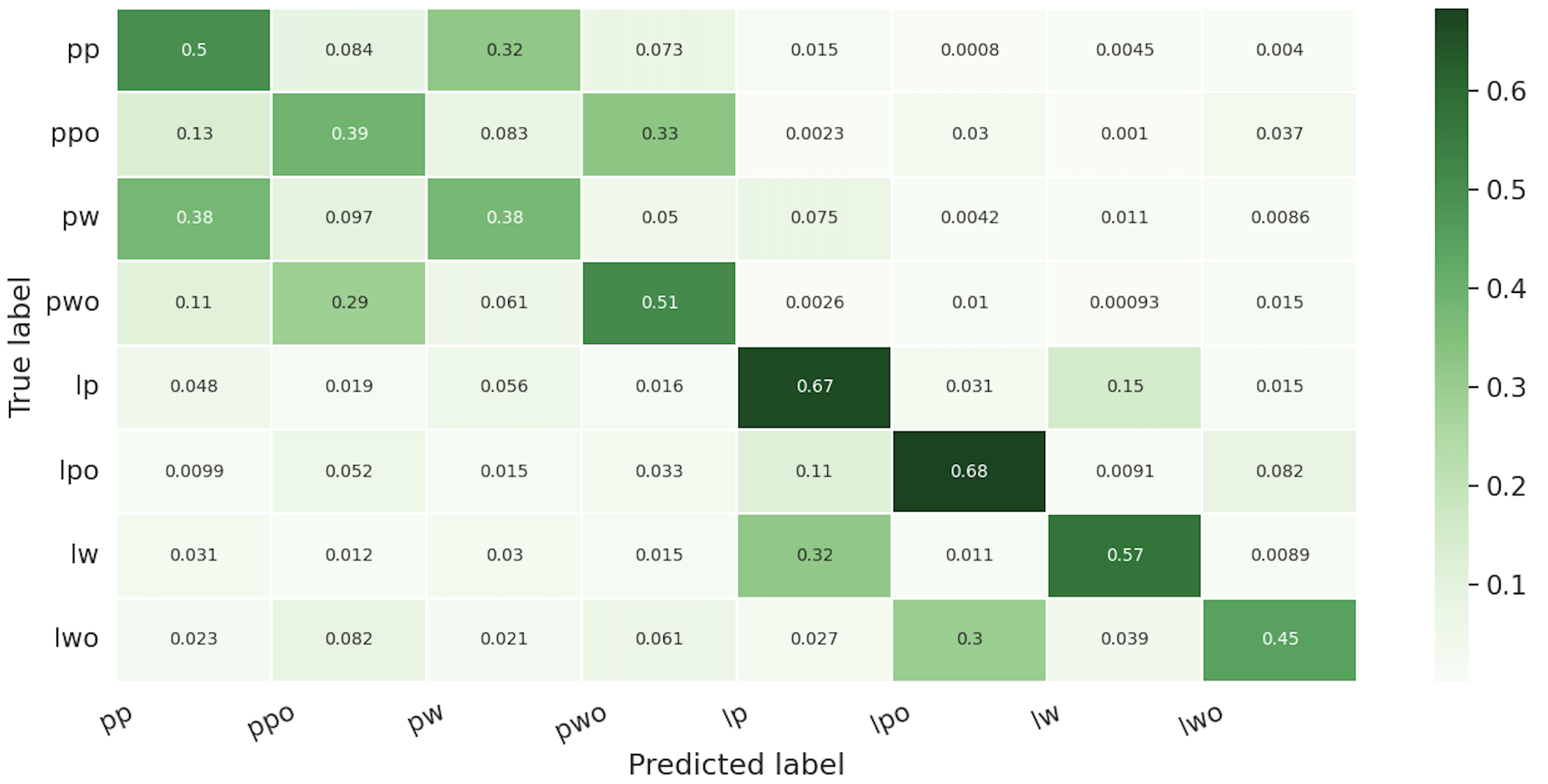}
    \caption{Confusion matrix for the random forest classifier trained to model the CHIME Catalog 2 data. The vertical axis is which model a given sample was from and the horizontal axis is how often each model was predicted to produce that sample.}
    \label{fig:confusionmatrix}
\end{figure*}
Training our random forest classifier yields a selection probability of $61.4\%$ for the LPO model, nearly five times higher than uniform random chance ($12.5\%$), indicating a clear preference for LPO over the alternative candidate models. To assess classifier accuracy, we generated confusion matrices using data simulated under each model's prior distribution (Fig.~\ref{fig:confusionmatrix}). The classifier reliably recovers the true underlying model in most cases, with misclassifications occurring primarily between models that differ only in their wait-time distribution (i.e.,~Poisson versus Weibull). The random forest classifier also ranks the LWO model as the second best fit to the CHIME data. If we assume that all models have equal probability of being selected, there is a 4.6\% chance that the two top models selected included a separate one off population for two separate methods of model selection. The chances of all four one-off models being preferred by one method (the learnt harmonic mean estimator) and the top two models being one-off models from a second method (the random forest classifier) is 0.31\%. This result suggests that the Catalog 2 data favors a subpopulation of truly one-off FRBs.

\subsection{Hyperactive FRBs}\label{sec:hyper}

We excluded the two hyperactive repeaters FRB~20180916B and FRB~20220912A from our fitting procedure, having 113 and 373 detections, respectively. However, we can use our posterior samples to test whether each model can naturally reproduce sources of this activity level. We define a source as hyperactive if it exceeds 35 detected bursts over the Catalog 2 timescale, which is the number of detections of the third most active FRB in the Catalog 2 dataset. For each model, we draw 2,000 posterior samples and record the number of hyperactive sources predicted in each simulation run. The resulting probability distributions are shown in Table~\ref{table:hyperactive_frbs}.

The results reveal a striking difference between models with and without a one-off subpopulation. The LPO model predicts the broadest range of hyperactive source counts, with a substantial probability of producing 5 or more hyperactive sources, consistent with the two outliers observed in Catalog 2. By contrast, models without a one-off subpopulation (LP, LW, PP, and PW) predict zero to five hyperactive sources in nearly all simulation runs, and in practice converge strongly on zero.

This difference can be traced directly to the inferred posterior distributions. As shown in Table~\ref{table:parameterfits}, models without a one-off subpopulation require a very large number of repeating sources with burst rates strongly skewed toward low values. This is the only way a purely repeating population can account for the large fraction of sources detected just once in Catalog 2. However, a population dominated by very low-rate repeaters is unlikely to produce the occasional hyperactive outlier. Models including a one-off subpopulation, by contrast, attribute the bulk of single-detection sources to genuinely one-off events, freeing the repeating population to occupy a broader range of burst rates and thus naturally accommodating hyperactive sources. That the LPO model reproduces both the overall burst count distribution and the existence of rare hyperactive outliers, without having been trained on the latter, provides an independent qualitative consistency check for the one-off model preference established by the Bayesian evidence and random forest classifier.

\begin{deluxetable}{l | C C C C}

\tablecaption{The probability distribution of the number of hyperactive repeating FRB sources predicted by each model. 
\label{table:hyperactive_frbs}}
\decimals
\tabletypesize{\normalsize}
\tablewidth{\textwidth}
\tablehead{\multirow{2}{*}[-2pt]{Model}& \multicolumn{4}{c}{\text{Number of Hyperactive Sources}} \\ \cline{2-5} & 0-5 & 5-10 & 10-15 & >15 }
\startdata
LPO & 0.16 & 0.39 & 0.27 & 0.18 \\
LWO & 0.28 & 0.51 & 0.18 & 0.03 \\
PPO & 0.93 & 0.07 & 0.0 & 0.0 \\
PWO & 0.98 & 0.02 & 0.0 & 0.0 \\
PW & 1.00 & 0.0 & 0.0 & 0.0 \\
PP & 1.00 & 0.0 & 0.0 & 0.0 \\
LW &  1.00 & 0.0 &0.0 & 0.0 \\
LP & 1.00 & 0.0 &0.0 & 0.0 \\
\enddata
\end{deluxetable}

\begin{deluxetable}{lrrrrrr}
\tablecaption{Evolution of the average observed percentage and number of CHIME 
sources seen to repeat ($N_{\rm rep}$), and the number of apparent non-repeaters 
($N_{\rm anr}$, intrinsically repeating sources with only one detection in the catalog), 
predicted by 2000 posterior samples of each model. All numbers  
exclude the first 6 months of Catalog 1, as detailed in 
Section \ref{chimesample}. For the bottom four models, where all sources by definition repeat,
we list $N_{\rm anr}$ simply as $\sim350$ and $\sim3500$ for Catalog 1 
and Catalog 2, respectively. See text for further details.
\label{table:forward_predict}}
\tabletypesize{\normalsize}
\tablewidth{\textwidth}
\tablehead{
  \colhead{} & 
  \multicolumn{3}{c}{Catalog 1} & 
  \multicolumn{3}{c}{Catalog 2} \\
  \colhead{Model} & 
  \colhead{(\%)} & 
  \colhead{$N_{\rm rep}$} &
  \colhead{$N_{\rm anr}$} &
  \colhead{(\%)} &
  \colhead{$N_{\rm rep}$} &
  \colhead{$N_{\rm anr}$}
}
\startdata
CHIME & \multicolumn{1}{c}{3.6} & \multicolumn{1}{c}{7}  & \multicolumn{1}{c}{14}  & \multicolumn{1}{c}{2.3} & \multicolumn{1}{c}{80}  & \multicolumn{1}{c}{---} \\
LPO   & $4.5 \pm 0.8$ & $17 \pm 7$  & $28 \pm 7$ & $2.4 \pm 0.4$ & $80 \pm 14$  & $80 \pm 25$ \\
LWO   & $5.4 \pm 1.2$ & $23 \pm 8$  & $48 \pm 18$ & $2.4 \pm 0.4$ & $79 \pm 14$  & $100 \pm 30$ \\
PPO   & $2.3 \pm 0.5$ & $8 \pm 4$   & $48 \pm 13$ & $2.6 \pm 0.6$ & $92 \pm 20$  & $250 \pm 90$ \\
PWO   & $1.9 \pm 0.1$ & $88 \pm 34$   & $80 \pm 40$ & $2.5 \pm 0.1$ & $446 \pm 210$  & $360 \pm 20$ \\
PW    & $0.5 \pm 0.1$ & $1\pm1$   & $\sim350$ & $0.1 \pm 0.0$ & $42 \pm 15$  & $\sim3500$ \\
PP    & $0.6 \pm 0.1$ & $1\pm1$   & $\sim350$ & $3.8 \pm 0.3$ & $62 \pm 9$   & $\sim3500$ \\
LW    & $6.9 \pm 1.4$ & $54 \pm 3$   & $\sim350$ & $14.8 \pm 1.0$ & $660 \pm 46$ & $\sim3500$ \\
LP    & $1.6 \pm 0.5$ & $5 \pm 2$   &$\sim350$ & $6.7 \pm 0.7$ & $197 \pm 20$ & $\sim3500$ \\
\enddata
\end{deluxetable}

\subsection{Volumetric rates}\label{sec:rates}
The constraints from our favored LPO model allow us to estimate volumetric rates for both repeating and one-off FRB sources observable to CHIME. We emphasize that these rates characterize the population detectable by the CHIME telescope out to a redshift of $z \sim 2$ \citep{Shin_2023}, rather than the intrinsic all-sky FRB population \citep{2024A&A...690A.377W}. Beaming and luminosity considerations would be required to extrapolate these estimates further.

For repeating FRBs, the posterior on the total number of sources $N$ is broad, spanning more than an order of magnitude in the 95\% credible interval (Table~\ref{table:parameterfits}). We therefore adopt a fiducial value of $N = 10^4$, consistent with the posterior median, and note that our rate scales linearly with this choice. Following \cite{Sautron_2025}, we adopt a magnetar lifetime of 5~Myr as a conservative upper limit, noting that no currently known magnetar exceeds this age. Combined with the comoving volume out to $z \sim 2$, this yields a volumetric rate of $\sim3$~Gpc$^{-3}$~Myr$^{-1}$ for repeating FRB sources. This volumetric rate is difficult to compare directly to Galactic magnetar birth rates \citep{Sautron_2025,2019MNRAS.487.1426B} without accounting for the number density of host galaxies and the fraction of magnetars that produce detectable FRBs. Accounting for both these quantities would require a luminosity function and beaming model that is beyond the scope of this work. Given the several orders of magnitude separating these rates, if repeating FRBs do originate from magnetars, only a small fraction of magnetars need produce detectable FRBs to reconcile the observations. We defer a detailed comparison to future analyses, but note that given the several orders of magnitude separating these rates, if repeating FRBs do originate from magnetars, only a small fraction of magnetars need produce detectable FRBs to reconcile the observations.

For one-off FRBs, the posterior is well-constrained, with a median rate of $r_{\rm oo} \sim 19$~FRBs~day$^{-1}$ in the CHIME sky. This translates to a volumetric rate of approximately 10~Gpc$^{-3}$~yr$^{-1}$, consistent with estimates of neutron star formation in globular clusters \citep{kremer_et_al}. However, this rate falls roughly three orders of magnitude below predictions for core-collapse and Type Ia supernovae \citep{Oguri_2019}, and approximately one order of magnitude below the compact object merger rate \citep{LIGO}.

Taken together, both repeating and one-off FRB rates fall significantly below the event rates of their most commonly proposed progenitors. This could indicate that FRBs accompany only a small fraction of these events, that emission is beamed and frequently directed away from us, or that CHIME's sensitivity limits it to detecting only a subset of FRBs produced by these populations. Distinguishing between these possibilities will require incorporating a luminosity function and beaming model into future analyses, as discussed in Section~\ref{sec:caveats}.

\subsection{Comparisons to other work}

Several previous analyses of the CHIME/FRB population have explored the repeating fraction of FRB sources \citep{clancy_frbs_can_repeat, amandas_repeater_paper, yamasaki2023truefractionrepeatingfast}. These works find results consistent with a wide range of intrinsic repeat fractions, from 50--100\%, and generally cannot rule out a solely repeating FRB population. While we do not rule out that all FRBs are capable of repeating in principle, we find that the Catalog 2 data decisively prefer a model in which a substantial fraction of sources appear as one-offs on observable timescales, whether due to intrinsically cataclysmic origins or to repetition timescales far exceeding current observing windows. Our predictions for the size of the repeating population are broadly consistent with \cite{yamasaki2023truefractionrepeatingfast}, though we infer a substantially higher rate of apparent one-off events. This difference likely reflects our explicit modeling of a one-off subpopulation, which those analyses did not include.

\subsection{Implications for future CHIME/FRB catalogs}\label{sec:implications}

Based on the favored LPO model, our 95\% credible interval sets an upper limit of approximately 36,000 repeating FRB sources and 19 one-off FRBs per day in the CHIME sky, both prior to selection effects. These limits suggest that the repeating population visible to CHIME is finite and will eventually saturate, while the one-off population continues to accumulate detections daily. 

Using the fitted parameters of each model, we estimate the rate $r_{\rm r}$ at which new repeaters would first appear in the CHIME sky, not accounting for a selection function. These values are listed in Table~\ref{table:parameterfits}. For the favored LPO model, $r_{\rm r} \sim 10$ sources per month. After accounting for selection effects, the actual rate of new sources seen by CHIME/FRB is about a tenth of this, and would require several millenia to reach saturation. Our models, therefore, predict no practical prospect of reaching a saturation point in the number of repeater detections.

\begin{figure}
    \centering
    \includegraphics[width=\linewidth]{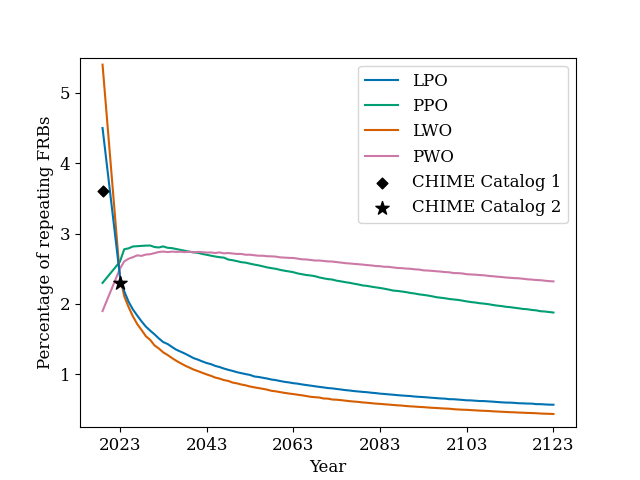}
    \caption{The evolution  of the repeater percentage of CHIME sources predicted by our top four models starting with the release of Catalog 1 and continuing for the 100 years following the release of Catalog 2.}
    \label{fig:forward_predic}
\end{figure}

We also use all of our models to determine the percentage and number of FRBs seen to repeat in Catalogs 1 and 2, as seen in Table~\ref{table:forward_predict}. The models including a subpopulation of one-offs far outperform those with a solely repeating population in predicting the observed repeater source percentages in both catalogs. The LPO model best fits both the percentage and number of repeating sources in each. While the LPO model is decisively favored, we show in Fig.~\ref{fig:forward_predic} the evolution of the repeater fraction for the top four models, tracking from Catalog 1 through the 100 years following Catalog 2. We extend the predictions over this timescale not because we expect CHIME to operate for a century, but to illustrate the long-term divergence in behavior between the various models. We derive these curves by taking the average repeater percentages of 2,000 posterior samples for each model at each timestamp.

All of these models predict a decline of the observed repeater source percentage seen by CHIME. The Poissonian burst-rate models predict a slight increase in the repeater fraction over the next few years, followed by a slight decline. These two models predict the FRB repeater percentage will remain approximately constant over the next century. The lognormal burst-rate models, on the other hand, predict a strong, consistent decline in the repeater fraction, with the evolution slowing as our repeater percentage reaches 0.5\%, as shown in Fig.~\ref{fig:forward_predic}. This means that, over time, we expect the CHIME sky to saturate with visible repeaters, while the daily one-offs continue to be detected by the telescope. In the short-term, assuming the next CHIME/FRB catalog comes out in the next five years, we would expect the repeater percentage to fall between 1.5\% and 2\%.

Also shown in Table~\ref{table:forward_predict} are the predicted numbers of apparent non-repeaters: sources that are intrinsically repeating in the model but produce only a single detectable burst during the catalog exposure period, making them observationally indistinguishable from truly one-off FRBs. The real CHIME data offer a natural calibration point: of the apparent non-repeaters in Catalog 1, 14 were subsequently observed to repeat by the time of Catalog 2. Across 2,000 posterior samples, our models with one-offs predict 80---250 such sources in the Catalog 2 timescale, with the LPO model predicting around 80. These results highlight that a substantial fraction of sources currently cataloged as one-offs will be revealed as repeaters with continued observation. In \cite{semisupervised_cat2_classification}, a semi-supervised machine learning method is used to classify repeaters and non-repeaters in CHIME Catalog 2 based on morphology. Their analyses predict approximately an order of magnitude more FRBs that may eventually repeat in Catalog 2 than the amount predicted by the LPO model and would be more in-line with the models in which all FRBs repeat (PW, PP, LW, LP). By definition, these models require that any non-repeating
source will ultimately repeat and are listed as $\sim 350$ and $\sim 3500$ in the respective columns for Catalog 1 and Catalog 2 where the numbers are chosen to simply represent the number of sources in each catalog. Our results, along with those of \cite{semisupervised_cat2_classification}, stand as testable predictions for future CHIME/FRB catalogs.

\subsection{Caveats and scope of this work}\label{sec:caveats}

The analyses presented in this work are deliberately simplified in several respects. Our primary goal is to establish SBI as a viable framework for modeling the CHIME/FRB population, and to demonstrate that meaningful population-level constraints can be extracted from a relatively minimal model. We view this work as a proof of concept, and discuss below how each simplification can be addressed in future analyses.

Our selection function models only the declination dependence of CHIME's sensitivity and exposure time. We do not incorporate the full selection function of \cite{kyle_selection_function}, which accounts for additional observational factors including dispersion measure, burst width, and spectral properties which was not available to us at the inception of this work. Incorporating the full selection function would substantially increase model complexity and computational cost, making it difficult to isolate the contribution of the SBI methodology itself. By showing that our simplified framework can already reproduce the Catalog 2 burst count distribution, match the time evolution of repeating and non-repeating sources, and predict hyperactive outliers without training on them, we demonstrate that SBI is a robust and flexible tool for this class of problem. Future work incorporating the full selection function of \cite{kyle_selection_function} will allow for more precise population constraints and will reduce the systematic uncertainties inherent in our current approach.

Similarly, we do not model a luminosity function or account for FRB beaming. As a consequence, our population constraints are explicitly observational, describing the population of FRBs detectable by CHIME rather than the intrinsic population of FRBs in the universe. The volumetric rates we derive in Section~\ref{sec:rates} should be interpreted with this in mind. Beaming and luminosity considerations would be required to map our constraints onto the full FRB population, and we defer this to future work. Importantly, however, the absence of these ingredients does not undermine our central result. The preference for a one-off subpopulation is driven by the shape of the burst count distribution including the tension between the large number of single-detection sources and the existence of hyperactive repeaters. This tension is present in the data regardless of the assumed luminosity function or beaming geometry. The SBI methodology we have developed is modular. The aforementioned simplifications can be incorporated in future analyses without requiring fundamental changes to the framework.

\section{Conclusion and Future Work}\label{sec:conc}
We have used SBI methods to model the potentially observable CHIME/FRB population. Our primary conclusions are as follows:

\begin{itemize}
    \item We developed a proof-of-concept framework showing the utility of SBI in modeling FRB populations. These methods will offer valuable insight to the true FRB population. Our methodology allowed us to very closely match the CHIME detections without assuming any luminosity, beaming, or morphological considerations.
    \item The CHIME/FRB sample decisively prefers underlying population models that have distinct one-off and repeating subpopulations. This preference is established independently by both the Bayesian evidence and a random forest classifier, with the latter selecting a one-off model as its top choice with a probability nearly five times greater than random chance.
    \item Our favored model (which employs a log-normal burst rate distribution with a Poissonian wait-time distribution and a subpopulation of one-off sources) is able to constrain the potentially observable repeating FRB population to 36,000 sources and the one-off rate to 19 FRBs/day. All models show that the potential for repeater discovery far outlasts the expected lifetime of CHIME.
    \item Our favored model predicts the fraction of CHIME sources seen to repeat will decline below 2\% within the next five years and asymptote toward 0.5\% on longer timescales as the CHIME sky saturates with one-off detections. The lognormal and Poissonian models diverge strongly in their long-term behavior, making the repeater fraction evolution a testable discriminant between them.
    \item Using our favored model, we calculate volumetric rates for one-off and repeating FRBs. Our volumetric rates fall below rate predictions for supernovae and magnetars, indicating that if these events are the cause of one-off and repeating FRBs, respectively, CHIME is not sensitive to all FRBs caused by these events.
    \item A substantial fraction of sources currently cataloged as apparent one-offs are predicted to be revealed as repeaters with continued observation, with the LPO model predicting $\sim$80 such sources in Catalog 2. These predictions, along with those from morphological studies by other authors, can be tested in future CHIME/FRB catalogs.
\end{itemize}

Future work will benefit from evolving this framework to include information on burst morphology, luminosity, and FRB beaming. Taking into account dispersion and other factors included in the full CHIME selection function \citep{kyle_selection_function} would be extremely useful in drawing further conclusions. With those additional considerations, these analyses could tell us more about the intrinsic population of FRBs. Testing these models on future CHIME/FRB catalogs will allow us to evaluate and improve upon the predictive power of the statistical methods presented here.

\begin{acknowledgments}
S.S. gratefully acknowledges the funding provided by the Ruby Foundation that supported this work under the Ruby Distinguished Doctoral Fellowship. D.R.L acknowledges support from the Eberly family. We thank Daniela Hoppenkothen and Jan Teusen for useful discussions. This work was supported by the National Science Foundation under award number 2606878.
\end{acknowledgments}

\begin{contribution}
    S.S.: conceptualization, methodology, software, formal analysis, investigation, visualization, writing -- original draft. D.R.L.: conceptualization, supervision, writing -- Review and editing.
\end{contribution}


\end{document}